\documentclass[useAMS,usenatbib,usegraphicx]{mn2e}
\usepackage{color}
\usepackage{mathptmx}

\newcommand{\aj}{{AJ}}
\newcommand{\apj}{{ApJ}}
\newcommand{\apjl}{{ApJ}}
\newcommand{\aap}{{A\&A}}
\newcommand{\aaps}{{A\&AS}}
\newcommand{\mnras}{{MNRAS}}
\newcommand{\pasa}{{PASA}}
\newcommand{\pasp}{{PASP}}

\title[UCD internal kinematics]{Spatially Resolved Kinematics of an Ultra-Compact Dwarf Galaxy\thanks{Based on observations obtained at the European Southern Observatory, Chile (Observing Programme 078.B-0496(B))}}
\author[M. J. Frank et al.]{M.~J.~Frank$^{1,2}$\thanks{E-mail: mfrank@ari.uni-heidelberg.de}, M.~Hilker$^{1}$, S.~Mieske$^{3}$, H.~Baumgardt$^{4}$, E.~K.~Grebel$^{2}$, and L. Infante$^{5}$\\
$^{1}$European Southern Observatory, Karl-Schwarzschild-Str. 2, 85748 Garching bei M\"unchen, Germany\\
$^{2}$Astronomisches Rechen-Institut, Zentrum f\"ur Astronomie der Universit\"at Heidelberg, M\"onchhofstr. 12-14, 69120 Heidelberg, Germany\\
$^{3}$European Southern Observatory, Alonso de Cordova 3107, Vitacura, Santiago, Chile\\
$^{4}$School of Mathematics and Physics, University of Queensland, QLD 4072, Australia\\
$^{5}$Departamento de Astronom\'ia y Astrof\'isica, Pontificia Universidad Cat\'olica de Chile, Casilla 306, Santiago 22, Chile\\
}
\begin{document}

\date{Accepted 2011 April 4.  Received 2011 March 30; in original form 2010 December 30}

\pagerange{\pageref{firstpage}--\pageref{lastpage}} \pubyear{2011}

\maketitle

\label{firstpage}

\begin{abstract}We present the internal kinematics of UCD3, the brightest known ultra-compact dwarf galaxy (UCD) in the Fornax cluster, making this the first UCD with spatially resolved spectroscopy. Our study is based on seeing-limited observations obtained with the ARGUS Integral Field Unit of the VLT/FLAMES spectrograph under excellent seeing conditions (0\farcs5 - 0\farcs67 FWHM). 

The velocity field of UCD3 shows the signature of weak rotation, comparable to that found in massive globular clusters. Its velocity dispersion profile is fully consistent with an isotropic velocity distribution and the assumption that mass follows the light distribution obtained from Hubble Space Telescope imaging. In particular, there is no evidence for the presence of an extended dark matter halo contributing a significant ($\ga$33 per cent within $R<200$~pc) mass fraction, nor for a central black hole more massive than $\sim$5 per cent of the UCD's mass.  
While this result does not exclude a galaxian origin for UCD3, we conclude that its internal kinematics are fully consistent with it being a massive star cluster.

\end{abstract}


\begin{figure*}
\includegraphics[width=0.57\textwidth]{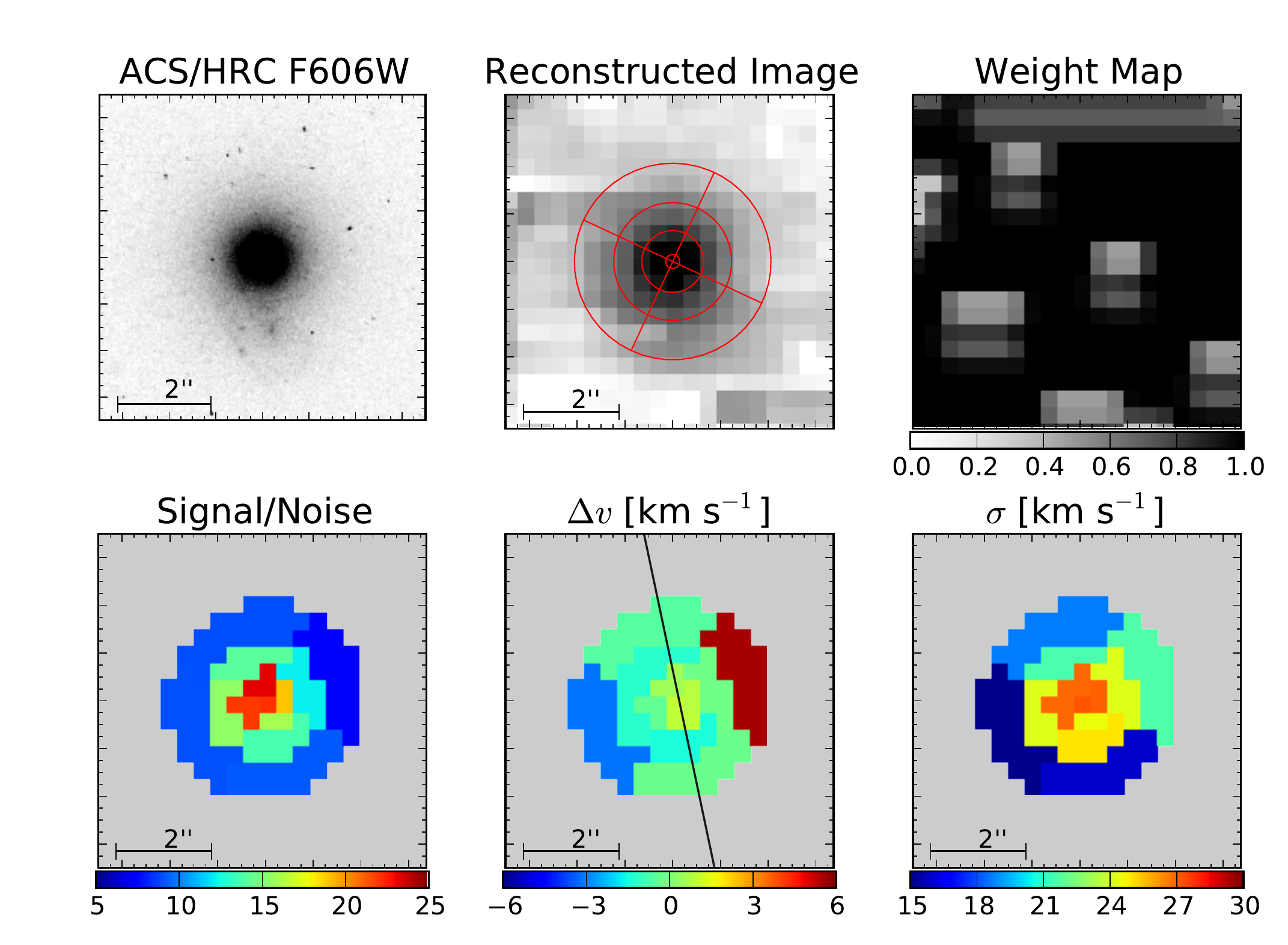}
\includegraphics[width=0.42\textwidth]{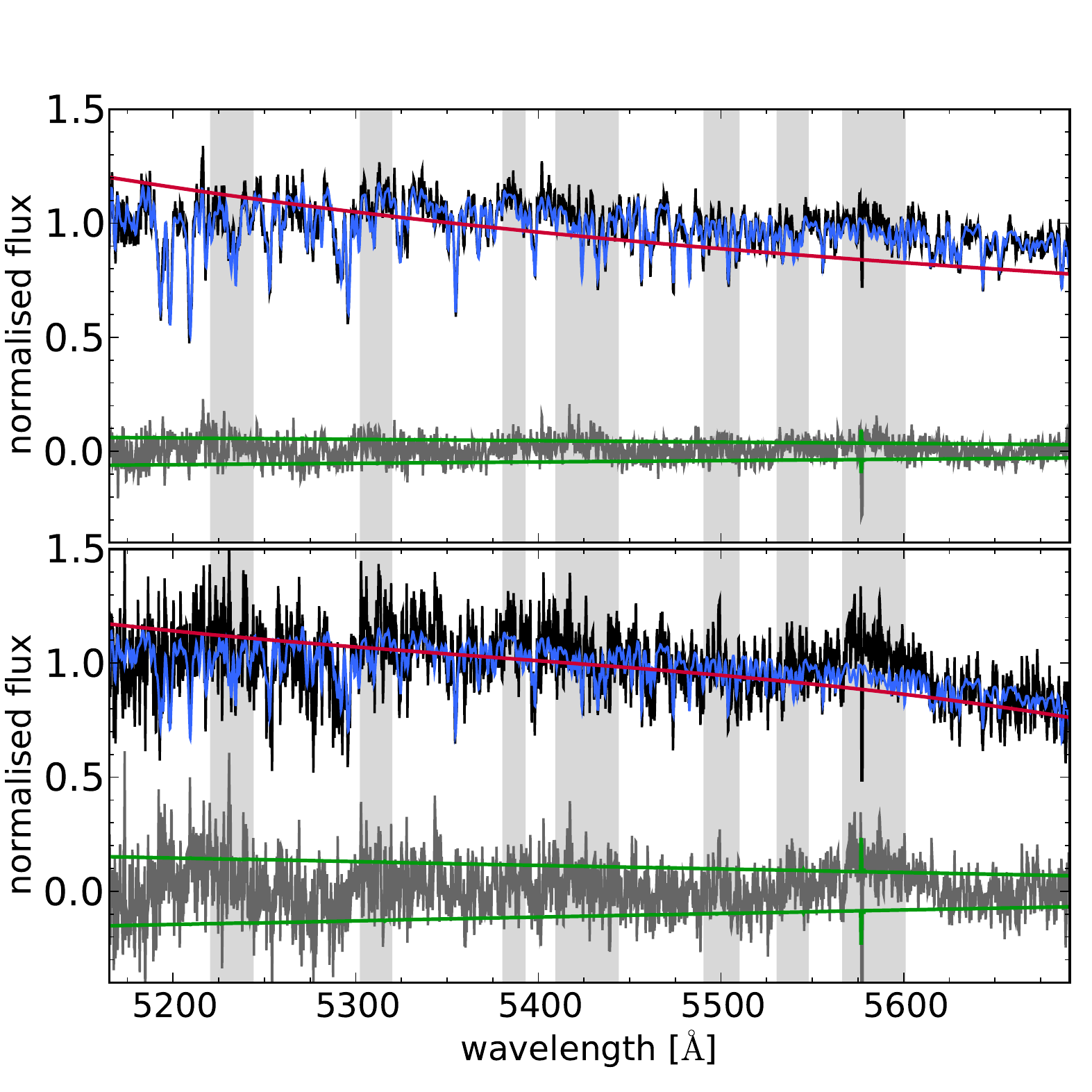}
\caption{Left: The upper row shows the ACS/HRC F606W image, the image reconstructed from the merged data cube with an overlaid polar grid used for the spatial binning and the weight-map, which illustrates the location of bad spaxels in the input data cubes. The bottom row shows the mean signal to noise ratio in each spatial bin per spectral bin (of 11\,km~s$^{-1}$ or $\sim0.2$\,\AA), the recovered velocity field with a line indicating the axis of rotation and the velocity dispersion map. The scale of the images is indicated by the $2\arcsec$ scale bar at the lower left, $1\arcsec$ corresponds to $\sim 92$\,pc at the distance of the Fornax cluster. Right: Two example spectra (black curves), the best-fitting spectral models (blue), residuals (grey) and noise models (green lines). The red line represents the third order multiplicative polynomial used in the fitting to absorb the continuum normalisation. The two spectra are representative for the range of quality in our data, corresponding to a central spatial bin (top) and an outer spatial bin, which is affected by the calibration lamp stray light (bottom).}
\label{figmaps}
\end{figure*}

\section{Introduction}
During the past decade, the existence of a previously unknown morphological class of compact stellar systems, called ultra-compact dwarf galaxies (UCDs), has been established. Objects of this type were first discovered in the dense environments of the nearby galaxy clusters like Fornax \citep{1999A&AS..134...75H,2000PASA...17..227D} and Virgo \citep{2005ApJ...627..203H}. They also were found around the dominant giant galaxies in nearby groups, e.g. NGC\,5128 \citep{2007A&A...469..147R,2010ApJ...712.1191T} and M\,104 \citep{2009MNRAS.394L..97H}, as well as in compact groups \citep{2011A&A...525A..86D}.

UCDs are characterised by predominantly old stellar populations \citep[$\ga\!8$ Gyr, e.g.][]{2011MNRAS.tmp...66C}, typical luminosities of $-14.0\!\la\! M_V\!\la\!-10.5$\,mag, half-light radii of $10\!\la\!r_h\!\la\!100$ pc \citep[e.g.][]{2007AJ....133.1722E} and central velocity dispersions of $20\!\la\!\sigma_0\!\la\!50$ km~s$^{-1}$, yielding dynamical masses of $2\times 10^6\!\la\! m\!\la\!10^8$ $\mathrm{M_{\odot}}$ \citep[e.g.][]{2007A&A...463..119H,2008A&A...487..921M}. Hence, they are larger, brighter and more massive than the most massive Milky Way globular clusters (GCs). At the same time they are significantly more compact than dwarf galaxies of comparable luminosity \citep{2008MNRAS.386..864D}.

One of the most striking ensemble properties of UCDs is that on average, their dynamical mass to light ratios (M/L) are $\sim$30-50 per cent higher than the predictions of canonical stellar population models \citep[e.g.][]{2009A&A...500..785K}. Moreover, their $V$-band M/L$_V$ are on average twice as large as those of Galactic GCs of comparable metallicity \citep[e.g.][]{2008A&A...487..921M,2008MNRAS.386..864D,2010ApJ...712.1191T}. Depending on their origin, this may indicate that UCDs mark the onset of dark matter domination in small stellar systems \citep{2008MNRAS.391..942B,2008MNRAS.385.2136G}, or harbour massive central black holes \citep{2009ApJ...699.1690M}, or exhibit an unusual initial stellar mass function \citep[][]{2008ApJ...677..276M,2009MNRAS.394.1529D,2010MNRAS.403.1054D}. 

The nature of UCDs is still uncertain. Do they represent the most massive star clusters \citep[][]{2002MNRAS.330..642F,2004A&A...418..445M}? Or are they the result of the environmental transformation of galaxies \citep[][]{2003MNRAS.344..399B}? 

Spatially resolved kinematics of UCDs open up a way to new insights on this matter, complementary to those gained from the study of the spatial and velocity distributions of UCD populations as a whole \citep[e.g.][]{2004A&A...418..445M}, of structural properties \citep{2005ApJ...623L.105D,2008AJ....136..461E}, of integrated internal velocity dispersions \citep[e.g.][]{2007A&A...463..119H}, or of their stellar population content \citep[e.g.][]{2010ApJ...712.1191T,2010ApJ...724L..64P}. Resolving the kinematics of objects with angular half-light radii on the order of $1\arcsec$, however, takes optical integral field spectroscopy to its limits. On the other hand, present day adaptive-optics aided, near infra-red integral field units (IFUs; e.g. Gemini/NIFS, \citealt{2003SPIE.4841.1581M}, VLT/SINFONI, \citealt{2003SPIE.4841.1548E}), provide only moderate spectral resolution ($\mathrm{R}=\Delta\lambda/\lambda\la4000$ or $\sigma_\mathrm{ins}\ga32$\,km~s$^{-1}$), limiting the ability to accurately sample the velocity distribution of typical UCDs with velocity dispersions of $\sim 25$\,km~s$^{-1}$.

In this Letter we present, for the first time, the spatially resolved kinematics of an UCD. A detailed description of the data reduction, of the technical aspects involved in the analysis and of the dynamical modelling will be presented in a forthcoming paper. The target of our study, UCD3, is one of the originally discovered objects \citep{1999A&AS..134...75H}, which subsequently led to the definition of the class of UCDs. Nevertheless, UCD3 is not an average UCD in several ways: With $M_{V}=-13.55$\,mag, it is the brightest known UCD in the Fornax cluster, separated by $\sim0.8$\,mag from the rest of the cluster's UCD population. In projection, it is relatively close to the central cluster galaxies NGC~1399 and NGC~1404 (projected distances of 46\,kpc and 13\,kpc, respectively). It is unusually red, with a spectroscopic metallicity of [Fe/H]$\sim-0.2$\,dex \citep{2008MNRAS.390..906C}. Most importantly, it has a large half-light radius of $r_\mathrm{h}=87$\,pc (or $\sim\,0\farcs95$) and is one of the few UCDs where an extended, low surface-brightness envelope has been detected \citep{2008AJ....136..461E}, making it possible to resolve it from the ground with seeing-limited instruments.

\section{Observations and data reduction}
\label{secobservations}
UCD3 was observed under excellent seeing with the ARGUS IFU \citep{2003Msngr.113...15K} of the Flames/GIRAFFE \citep{2002Msngr.110....1P} spectrograph mounted on UT2 at the VLT. The IFU was used in the 1:1 scale, yielding a field of view of $11\farcs 5\times 7\farcs 3$ at a spatial sampling of $0\farcs52\times 0\farcs52$ per lenslet (or spaxel), with the LR04 grism, providing a spectral resolution of R$\!\sim\!9600$ (or $\sim\!13$\,km~s$^{-1}$ in terms of Gaussian $\sigma$) over the range $5015\!-\!5831$~\AA. The observations were executed in service mode over the period of almost one year and consisted of eight exposures of $2775$ seconds each. 

The basic data reduction and the extraction of wavelength calibrated spectra was carried out using the Giraffe Baseline Data Reduction System \citep{2000SPIE.4008..467B}, standard IRAF tasks and custom python routines. The sky spectrum was estimated from IFU spectra outside an aperture of $r\ge7$ spaxels around the object's centre and the dedicated sky fibres placed around the IFU. The data were corrected for heliocentric velocities and the observations from all nights were resampled to a common logarithmic wavelength scale and organised into 3d data cubes. 
For each data cube, the centre of the UCD and the effective seeing were determined by fitting a model image to the reconstructed image. The model image consisted of an ACS high resolution channel (HRC) image in the F606W filter \citep[shown in the upper left panel of Fig.~\ref{figmaps}; HST programme 10137,][]{2007AJ....133.1722E}, convolved with a Gaussian PSF and rebinned to the IFU's spatial scale. The five observations with the best seeing ($0\farcs50\le$ PSF FWHM $\le0\farcs67$, with a mean of $0\farcs60$), were then stacked by resampling them to a common coordinate grid, at the same time removing the effect of atmospheric dispersion \citep{1982PASP...94..715F}. The resampling was done using the drizzle algorithm \citep{2002PASP..114..144F}, treating each spatial slice as an image and masking bad spaxels (cf. the weight map in Fig.~\ref{figmaps}). These comprised a hot column on the detector (see the Flames User Manual), as well as spectra that were strongly affected by the stray light from neighbouring simultaneous calibration lamp spectra. To minimise the loss of spatial resolution due to resampling, a moderate amount of super-sampling was applied by using output pixels smaller by a factor of 1.5 on each side. The corresponding reconstructed image is shown in Fig.~\ref{figmaps}.

The spatial extent of the UCD, with a half-light radius of $0\farcs95$, is only $\sim$2-3 times as large as the spaxel size and the seeing HWHM. In order to preserve the structure of the object, while achieving a higher signal to noise ratio in the outer parts of the UCD, we used a geometrical scheme for binning the limited number of spaxels receiving significant flux from the UCD: The spaxels were grouped into radial annuli, which were subdivided into quadrants (shown as red polar grid overlaid on the reconstructed image in Fig.~\ref{figmaps}).

\section{Kinematic Measurements}
\label{secLOSVDs}
The line-of-sight velocity $v$ and velocity dispersion $\sigma$ were recovered from each spectrum using the penalised pixel-fitting (ppxf) code of \citet{2004PASP..116..138C}. Two exemplary spectra with their best-fitting models are shown on the right of Fig.~\ref{figmaps}.
As spectral templates for the fitting we used spectra of field stars and members of the NGC\,6475 open cluster from the UVES Paranal Observatory Project \citep[UVESPOP,][]{2003Msngr.114...10B} library. The library consists of reduced and calibrated high-resolution ($R\!\sim\!80000$) stellar spectra spanning a range of spectral-types and metallicities. We used a sub-sample of 46 late-type stars (spectral types F0 to M1, luminosity classes II-V), of which the ppxf algorithm would select a subset of $\la\!8$ significant templates during the fitting. The spectra were brought to a common radial velocity and rebinned to the same logarithmic wavelength-scale as our ARGUS data. To degrade the template library to the resolution of our observations, we constructed an empirical model of the spectral line-spread function (LSF) as a function of wavelength \citep[e.g.][]{2007MNRAS.376.1033C}, by fitting UVESPOP templates of the according spectral type to late-type stars observed with the ARGUS LR04 setup, and convolved the templates with this LSF.

The magnitude of the noise in our spectra varied gradually with wavelength. Therefore it was modelled by using the DER\_SNR algorithm \citep{2007STECF..42....4S} to estimate the noise at each wavelength and then fitting a linear function to these estimates. The resulting noise spectrum (shown as green lines in the spectra in Fig.~\ref{figmaps}) was used as an input for ppxf to ensure the correct relative weighting of pixels in the $\chi^2$-minimisation, and as a basis for generating 100 Monte Carlo realisations of each spectrum, which were used to obtain uncertainties on the measured velocity and velocity dispersion.

The fitting was done in the wavelength range from $5156$\,\AA\ to $5691$\,\AA. We masked out spectral regions containing a strong emission line doublet in some of the spaxels, most likely originating from a background galaxy slightly offset from the UCD's position (cf. 
\citealt{2007AJ....133.1722E}), the 5577\,\AA\ O{\scshape{I}} sky line, as well as regions particularly affected by the calibration lamp stray light. For consistency the same mask (represented by the grey wavelength intervals in Fig.~\ref{figmaps}) was applied to all spectra, regardless of whether they were affected by these blemishes or not.

\section{Internal Dynamics}
\label{secinternaldynamics}
\subsection{Rotation}
\label{secrotation}
\begin{figure}
\includegraphics[width=0.49\textwidth]{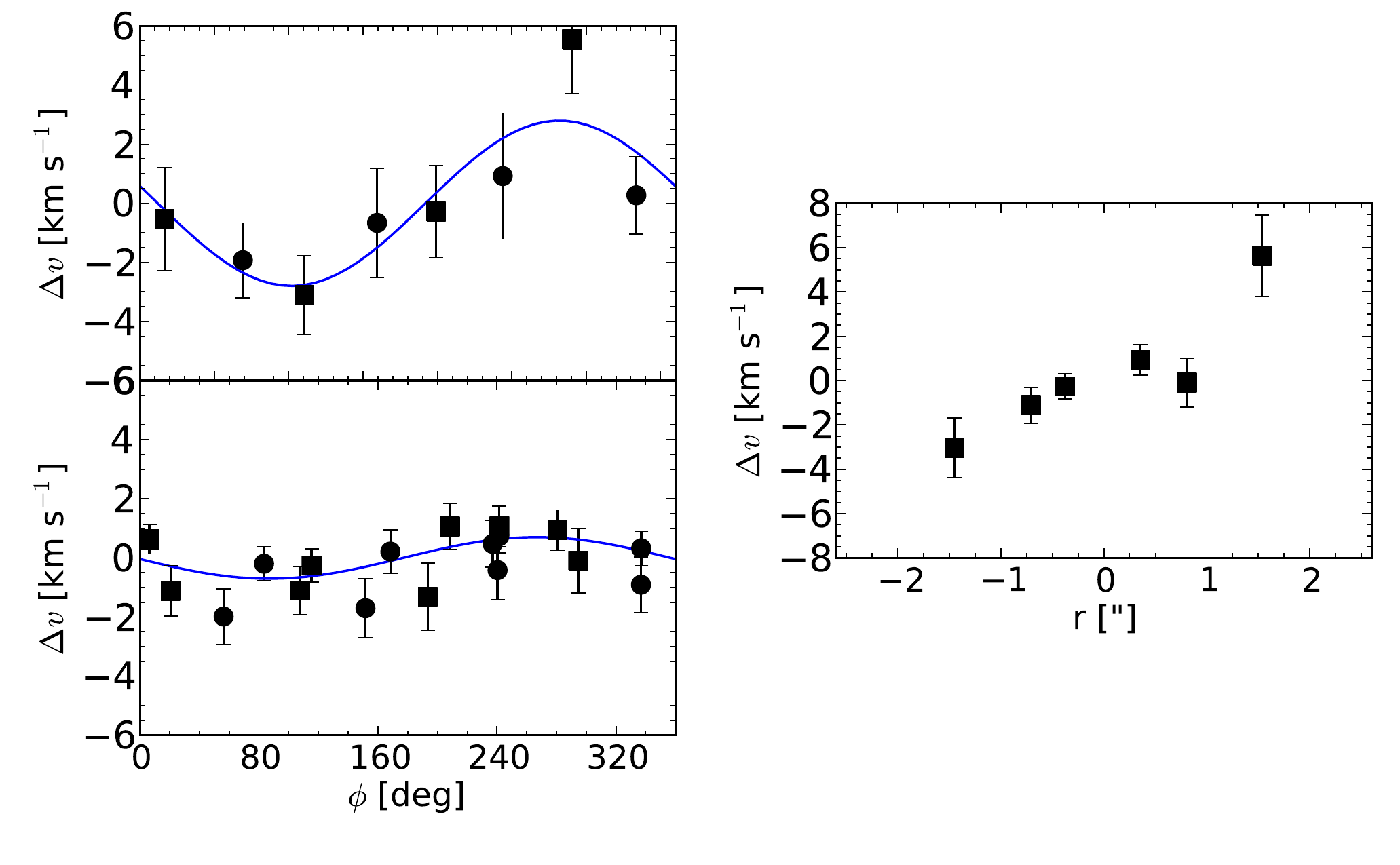}
\caption{The signature of rotation in the UCD. Left: The top panel shows the measured velocity versus the luminosity-weighted azimuth $\phi$ of each bin in the outer annulus (r$\ge1\farcs2$) of the spatial binning shown in Fig.~\ref{figmaps} (squares) and the same binning with the quadrants rotated by $45$ degrees (circles). While the orientation of the quadrants in Fig.~\ref{figmaps} is chosen such that the maximum rotational signal is captured, the signal essentially disappears when the quadrants are rotated by $45$ degrees. The blue curve represents the best-fitting sinusoid (see Sect.\ \ref{secrotation}) to the velocities measured in both orientations of the quadrants (i.e. square and circle symbols). The bottom panel shows the same for the inner annuli (r$<1\farcs2$). Right: The rotation curve of the UCD along the direction approximately perpendicular to the kinematic axis.}
\label{figrotationazimuthalradial}
\end{figure} 

The resulting velocity field of the UCD shows the signature of weak rotation (see the second panel in the bottom row of Fig.~\ref{figmaps}). To recover the projected kinematic axis, we initially varied the position angle of the quadrants in the spatial binning; once its approximate orientation was found, a sinusoid of the form $v(r)=v_{0}+A\,\mathrm{sin}\left(\phi-\textrm{PA}\right)$ was fitted simultaneously to the velocities recovered from that orientation and from the quadrants rotated by $45$ degrees (Fig.~\ref{figrotationazimuthalradial}). Here $v_{0}$ is the systemic velocity, $A$ is the amplitude of the rotation, $\phi$ is the azimuth of each bin, and $\textrm{PA}$ is the position angle of the rotational axis. In the outer annulus (r$\ge1\farcs2$; upper left panel of Fig.~\ref{figrotationazimuthalradial}), the fit yielded a $\textrm{PA}$ of $12\pm15$ degrees and an amplitude of $2.8\pm0.7$\,km~s$^{-1}$. Due to the unknown inclination of the rotational axis and due to the fact that the maximum of the rotation curve velocity need not be reached at r$\sim1\farcs5$, this represents a lower limit on the true rotational velocity.

\subsection{Dispersion profile}
\begin{figure}
\includegraphics[width=0.49\textwidth]{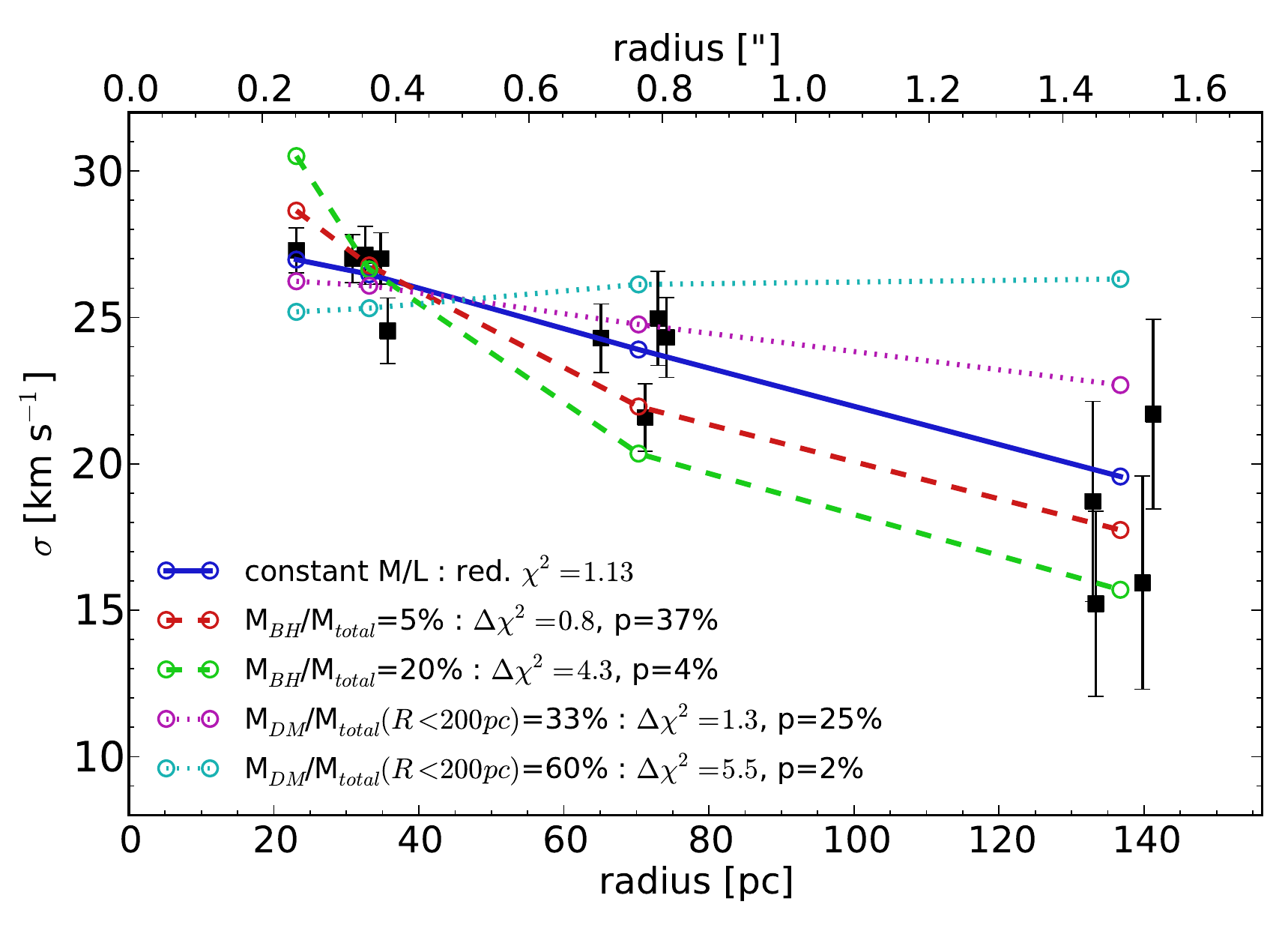}
\caption{The measured velocity dispersions are shown as black squares. The radii of the data points correspond to the luminosity-weighted radius of each spatial bin. Models based on an isotropic velocity distribution are represented by open circles (evaluated in each radial annulus), connected by curves to guide the eye: A `mass follows light' model (blue, solid curve), two models including a massive black hole of 5 (red) and 20 per cent (green, dashed curves) of the UCD's mass and two models including a dark matter component contributing 33 (magenta) and 60 per cent (cyan, dotted curves) of the total mass within a three-dimensional radius of 200\,pc.}
\label{figdispprofile}
\end{figure}
The radial dispersion profile of UCD3 is shown in Fig.~\ref{figdispprofile}. The dispersion decreases continuously with increasing radius. There is no sign of a disturbed profile, which could have been expected if the UCD had undergone a recent interaction, leaving it out of dynamical equilibrium, a scenario proposed to explain the high dynamical masses of some UCDs \citep{2006MNRAS.367.1577F}.

Along with the observed velocity dispersion profile, Fig.~\ref{figdispprofile} shows a number of models. The models are based on the best-fitting two-component luminosity distribution determined by \citet{2007AJ....133.1722E}, a Fornax cluster distance of 18.97\,Mpc \citep{2001ApJ...553...47F} and an isotropic velocity distribution. They were constructed using the code described in \citet{2007A&A...463..119H} and adapted to integral-field data. The code takes into account the convolution with the seeing PSF and predicts the full line-of-sight velocity distribution (LOSVD) in each of our radial annuli. While the measurement of the observed velocity dispersion assumed a Gaussian LOSVD, the actual LOSVD need not be strictly Gaussian. In the case of the models including a massive black hole (Sect.~\ref{secbhmodels}), where a significant fraction of the observed stars move in the potential of the central point source, the integrated velocity distribution can even become highly non-Gaussian \citep[cf.][]{2009ApJ...699.1690M}. Therefore, to obtain the model velocity dispersions that correspond to the full modelled LOSVDs produced by the model, we created artificial spectra by convolving UVESPOP spectra with the full LOSVDs and then measured the dispersion in the same way as for the observed spectra. 

Using the fact that the velocity dispersion $\sigma$ scales with the mass as $\sigma^2\propto\textrm{M}$ (virial theorem), the best-fitting mass (or equivalently the best-fitting M/L, for a given observed luminosity L) for each model was calculated by scaling the model to the data \citep[eq.~51 in][]{2008MNRAS.390...71C}, so that $\hat\chi^2=\sum_i{(\sigma_\mathrm{m,i}-\sigma_\mathrm{obs,i})^2/(\Delta_{\mathrm{obs,i}}^2)}$ is minimised, where  $\sigma_\mathrm{obs,i}$ are the observed velocity dispersions, $\Delta_{\mathrm{obs,i}}$ are their associated uncertainties and $\sigma_\mathrm{obs,i}$ are the model velocity dispersions at the same radii. Confidence regions around the best-fitting parameters were estimated from the increase in the reduced $\chi^2 = k^{-1}\hat\chi^2$ compared to its minimum value, where $k = (\textrm{number of data points}-\textrm{number of model parameters}-1)$, with the number of model parameters being 1 for the constant M/L model and 2 for the models including either dark matter or a black hole. For one parameter of interest, the 68, 90 and 99 percent confidence regions are given by an increase $\Delta\chi^2 = \chi^2 - \chi^2_\mathrm{min}$ of 1, 2.71 and 6.63 \citep[e.g.][]{UBHD-66547151}. For an arbitrary $\Delta\chi^2$, the confidence level can be found by numerical integration of the $\chi^2$ distribution function. The minimum reduced $\chi^2$ for the best-fitting model, as well as the $\Delta\chi^2$ for the other models along with the probabilities to find the true model parameter at this $\Delta\chi^2$ are given in the legend of Fig.~\ref{figdispprofile}.

The simplest possible model, assuming that mass follows light (or equivalently, a constant M/L; shown as blue, solid curve) provides an excellent fit to the observed dispersion profile, implying a UCD mass of M=$8.2\pm0.7\times10^{7}$\,M$\odot$ and a M/L ratio of M/L$_{V}$=$3.6\pm0.3$. The uncertainty corresponds to the 1-$\sigma$ confidence interval in the sense of the previous paragraph, obtained by varying M/L until $\chi^2$ increased by $\Delta\chi^2 = 1$. The M/L is in good agreement with the stellar population M/L$_{V}$ of $3.7\pm0.2$ \citep{2011MNRAS.tmp...66C}. However, regarding the stellar population M/L of UCD3, we note that there is a disagreement on the age of UCD3 in the literature: Lick/IDS absorption indices suggest an age of $\sim2-5$\,Gyr \citep{2006AJ....131.2442M,2009MNRAS.394.1801F}. This would render UCD3 one of the youngest known UCDs and imply a stellar  M/L$_{V}\la2$, which in turn would still require substantial dark mass according to our measured dynamical M/L ratio. On the other hand, studies based on spectral fitting yield ages of $\ga12$\,Gyr \citep{2008MNRAS.390..906C,2011MNRAS.tmp...66C}, placing UCD3 within the typical range of ages found for most UCDs, which would make our dynamical M/L ratio fully consistent with a standard stellar mass function. As a possible cause for the discrepant ages, \citet{2009MNRAS.394.1801F} point out that the young age inferred from absorption indices could be an artifact introduced by the presence of hot horizontal branch stars.

\subsection{Constraints on the non-luminous matter content}
\subsubsection{Dark matter}
As a possible explanation for the, on average, high dynamical mass to light ratios of UCDs, a substantial dark matter content is discussed in the literature \citep[e.g.][]{2005ApJ...627..203H,2008MNRAS.391..942B}. A potential formation channel would then be the tidal stripping of nucleated galaxies, leaving behind a remnant nucleus with properties comparable to those of a UCD \citep{2003MNRAS.344..399B}. Simulations show that the resulting UCDs can be dark matter dominated, if the central concentration of dark matter was enhanced prior to the stripping through, for example, the adiabatic infall of gas \citep{2008MNRAS.385.2136G,2008MNRAS.391..942B}.

In order to test this scenario, we included in the mass model a dark matter component with a density profile mimicking the dark matter distribution predicted for the final evolutionary stage by the simulations of \citet{2008MNRAS.385.2136G}. Their Fig.~5 shows a final dark matter density approximately proportional to $r^{-2.5}$ for radii $200$\,pc$\la r\la2\,$kpc. We chose as an analytic expression in our model a profile of the form $\rho = \rho_0 (1+r^2/r_s^2)^{-1.25}$, which shows the same slope at large radii and converges to a central density $\rho_0$ at $r=0$. As a scale radius we adopted $r_s=200$\,pc being the approximate radius below which the \citet{2008MNRAS.385.2136G} profile starts to flatten. We calculated models for a fraction of 20, 33, 60 and 67 per cent of dark matter within a three-dimensional radius of 200\,pc. We find that, while yielding a worse fit than the model without dark matter, a dark matter fraction of 33 per cent (shown as magenta dotted curve in Fig.~\ref{figdispprofile}) is, with $\Delta\chi^2=1.3$ (again for one parameter of interest, we marginalise over the M/L), compatible at the 1-$\sigma$ level with the observed velocity dispersion profile and yields a stellar M/L$_{V}$ of 3.4. A model with a dark matter fraction of 60 per cent (cyan dotted curve) and a best-fitting stellar M/L$_{V}$ of 2.7, is excluded by the data with 98 per cent confidence ($\Delta\chi^2=5.5$).

\subsubsection{Massive black hole}
\label{secbhmodels}
Unseen mass in UCDs could also be present in the form of a massive black hole. In massive galaxies, nuclear star clusters can coexist with central black holes with comparable masses \citep{2009MNRAS.397.2148G}. Therefore, if UCDs represent the remnant nuclei of originally much more massive galaxies, at least some of them could contain a black hole which contributes a substantial fraction of the total mass. Also if UCDs represent compact star clusters around recoiling super-massive black holes that were ejected from the centres of massive galaxies \citep{2009ApJ...699.1690M}, they would contain a black hole with a mass comparable to that of the stellar component. Therefore, we explored a family of mass models which, in addition to the constant M/L stellar component, included massive black holes with 5, 10, 15 and 20 per cent of the UCDs total mass. 

The resulting model dispersion profiles, for a black hole mass of 5 (red) and 20 per cent (green) are shown as dashed curves in Fig.~\ref{figdispprofile}. The models yield best-fitting stellar M/L$_{V}$ of 2.7 (M$_\mathrm{BH}$/M$_\mathrm{total}$=5 per cent) and 1.8 (M$_\mathrm{BH}$/M$_\mathrm{total}$=20 per cent) and provide a worse fit to the data than the model without a black hole, with M$_\mathrm{BH}$/M$_\mathrm{total}=$5 per cent still being consistent with the data at the 1-$\sigma$ level ($\Delta\chi^2=0.8$), and the M$_\mathrm{BH}$/M$_\mathrm{total}=$20 per cent model being excluded with 96 per cent confidence ($\Delta\chi^2=4.3$).

\section{Conclusions}
We have demonstrated that the resolution of the internal kinematics of the brightest and most extended UCDs is feasible with seeing-limited integral-field spectroscopy, making it a promising tool to constrain the formation scenarios and hence the nature of UCDs.  Our analysis of UCD3 in the Fornax cluster reveals weak rotation and a dispersion profile in excellent agreement with an isotropic velocity distribution and a constant M/L. The mass of the UCD, measured under these assumptions, is $8.2\pm0.7\times10^{7}$\,M$\odot$ and the implied M/L is in good agreement with the most recent determination of the UCD's stellar population age and metallicity. The UCD does not show any of the attributes, such as tidal disturbance, a black hole with a mass comparable to that of the UCD, or a significant dark component, that would strongly suggest that the UCD is the remnant nucleus of a larger galaxy or a remnant star cluster around a recoiling super-massive black hole. While this result does not exclude such scenarios, the internal kinematics of UCD3 are fully consistent with it being a massive globular cluster.

\section*{Acknowledgements}
We thank the anonymous reviewer for valuable comments. MJF would like to thank Marina Rejkuba, Eric Emsellem and Harald Kuntschner for helpful discussions about this project. HB acknowledges support from the Australian Research Council through Future Fellowship grant FT0991052.

\bsp

\label{lastpage}

\end{document}